\documentclass[12pt]{iopart}

\usepackage{iopams}  

\newcommand{\beq}{\begin{equation}}
\newcommand{\eeq}{\end{equation}}
\newcommand{\beqs}{\begin{eqnarray}}
\newcommand{\eeqs}{\end{eqnarray}}

\begin{document}

\title{Some Exact Results for Spanning Trees on Lattices} 

\author{Shu-Chiuan Chang$^{a}$ and Robert Shrock$^{b}$}

\address{(a) \ Department of Physics \\
National Cheng Kung University \\
Tainan 70101, Taiwan}

\address{(b) \ C. N. Yang Institute for Theoretical Physics \\
State University of New York \\
Stony Brook, N. Y. 11794 \\
USA}

\begin{abstract}

For $n$-vertex, $d$-dimensional lattices $\Lambda$ with $d \ge 2$, the number
of spanning trees $N_{ST}(\Lambda)$ grows asymptotically as $\exp(n z_\Lambda)$
in the thermodynamic limit. We present an exact closed-form result for the
asymptotic growth constant $z_{bcc(d)}$ for spanning trees on the
$d$-dimensional body-centered cubic lattice. We also give an exact integral
expression for $z_{fcc}$ on the face-centered cubic lattice and an exact
closed-form expression for $z_{488}$ on the $4 \cdot 8 \cdot 8$ lattice.

\end{abstract}

\maketitle

\section{Introduction}

 Let $G=(V,E)$ denote a connected graph (without loops) with vertex (site) and
edge (bond) sets $V$ and $E$.  Let $n=v(G)=|V|$ be the number of vertices and
$e(G)=|E|$ the number of edges in $G$.  A spanning subgraph $G^\prime$ is a
subgraph of $G$ with $v(G^\prime) = |V|$, and a tree is a connected subgraph
with no circuits.  A spanning tree is a spanning subgraph of $G$ that is a tree
(and hence $e(G') = n-1$).  A problem of fundamental interest in mathematics
and physics is the enumeration of the number of spanning trees on the graph
$G$, $N_{ST}(G)$.  This number can be calculated in several ways, including as
a determinant of the Laplacian matrix of $G$ and as a special case of the Tutte
polynomial of $G$ \cite{fh,bbook}.  In this paper we shall present an exact
closed-form result for the asymptotic growth constant for spanning trees on the
$d$-dimensional body-centered cubic lattice, denoted $bcc(d)$, with $bcc(3)
\equiv bcc$.  We shall also give an exact integral expression for the
$z_{fcc}$ describing the face-centered cubic lattice and an exact 
closed-form expression $z_{488}$ for the $4 \cdot 8 \cdot 8$
lattice.  A previous study on the enumeration of spanning trees and the
calculation of their asymptotic growth constants was carried out in
Ref. \cite{sw}. In that work, closed-form integrals for these quantities were
given, and from the integral for the $bcc(d)$ lattice, an infinite series
representation was derived.  Our present result for the $bcc(d)$ lattice is
obtained by summing exactly this infinite series.  Similarly, our present
result for the $4 \cdot 8 \cdot 8$ lattice is obtained by an exact closed-form
evaluation of the integral given for this lattice in Ref. \cite{sw}.

\section{Background and Method}

We briefly recall some definitions and background on spanning trees and the
calculational method that we use.  For $G=G(V,E)$, the degree $k_i$ of a vertex
$v_i \in V$ is the number of edges attached to it. A $k$-regular graph is a
graph with the property that each of its vertices has the same degree $k$.  Two
vertices are adjacent if they are connected by an edge. The adjacency matrix
$A(G)$ of $G$ is the $n \times n$ matrix with elements $A_{ij}=1$ if $v_i$ and
$v_j$ are adjacent and zero otherwise. The Laplacian matrix $Q=Q(G)$ is the $n
\times n$ matrix $Q$ with $Q_{ij}=k_i\delta_{ij}-A_{ij}$.  One of the
eigenvalues of $Q(G)$ is always zero; let us denote the rest as $\lambda_i(G)$,
$1 \le i \le n-1$.  A basic theorem is that \cite{fh,bbook} $N_{ST}(G) =
(1/n)\prod_{i=1}^{n-1} \lambda_i(G)$.  Here we shall focus on $k$-regular
$d$-dimensional lattices $\Lambda$.  For these lattices, if $d \ge 2$, then in
the thermodynamic limit, $N_{ST}$ grows exponentially with $n$ as 
$n \to \infty$; that is, there exists a constant $z_\Lambda$ such that 
$N_{ST}(\Lambda) \sim \exp(n z_\Lambda)$ as $n \to \infty$.  The constant 
describing this exponential growth is thus given by 
\beq
z_{\Lambda} = \lim_{n \to \infty} n^{-1} \ln N_{ST}(\Lambda) \ . 
\label{zdef}
\eeq
where $\Lambda$, when used as a subscript in this manner, implicitly refers to
the thermodynamic limit of the lattice $\Lambda$.  A regular $d$-dimensional
lattice is comprised of repeated unit cells, each containing $\nu$ vertices.
Define $a(\tilde n,\tilde n')$ as the $\nu \times \nu$ matrix describing the
adjacency of the ($d$-dimensional) vertices of the unit cells $\tilde n$ and
$\tilde n'$, the elements of which are given by $a(\tilde n,\tilde n')_{ij}=1$
if $v_i \in \tilde n$ is adjacent to $v_j \in \tilde n'$ and 0 otherwise.
Assuming that a given lattice has periodic boundary conditions, and using the
resultant translational symmetry, we have $a(\tilde n, \tilde n')= a(\tilde n-
\tilde n')$, and we can therefore write $a(\tilde n)=a(\tilde n_1,\cdots,\tilde
n_d)$.  In Ref. \cite{sw} a method was derived to calculate $N_{ST}(\Lambda)$
and $z_{\Lambda}$ in terms of a matrix $M$ which is determined by these
$a(\tilde n, \tilde n')$.  For a $d$-dimensional lattice, define
\beq
M(\theta_1,\cdots,\theta_d) = k \cdot 1 - \sum_{\tilde n} 
a(\tilde n) e^{i \tilde n \cdot \theta } 
\label{mmatrix}
\eeq
where in this equation $1$ is the unit matrix and $\theta$ stands for the
$d$-dimensional vector $(\theta_1,\cdots,\theta_d)$.  Then \cite{sw} 
\beq
z_{\Lambda} =  {1\over \nu }\int_{-\pi}^\pi 
\biggl [ \prod_{j=1}^d {d\theta_j \over {2\pi}} \biggr ] 
\ln[{\rm det}(M(\theta_1,\cdots,\theta_d))] 
\label{zint}
\eeq

For a $k$-regular graph $\Lambda$, a general upper bound is $z_{\Lambda} \le
\ln k$.  A stronger upper bound for a $k$-regular graph $\Lambda$ with
coordination number $k \ge 3$ can be obtained from the bound
\cite{mckay,chungyau}
\beq
N_{ST}(G) \le \Biggl ( \frac{2\ln n}{n k \ln k} \Bigg) (C_k)^n
\label{nmckay}
\eeq
where
\beq
C_k = \frac{(k-1)^{k-1}}{[k(k-2)]^{\frac{k}{2}-1}} \ . 
\label{ck}
\eeq
With eq. (\ref{zdef}), this then yields \cite{sw} 
\beq
z_{\Lambda} \leq \ln(C_k) \ . 
\label{mcybound}
\eeq
It is of interest to see how close the exact results are to these upper
bounds.  For this purpose, we define the ratio
\beq
r_{\Lambda} = \frac{z_{\Lambda}}{\ln C_k} 
\label{rupper}
\eeq
where $k$ is the coordination number of $\Lambda$. 

\section{$bcc(d)$ Lattice}

For the $bcc(d)$ lattice a unit cell contains $\nu_{bcc(d)}=2$ vertices located
at $v_1= (0,\cdots,0)$ and $v_2 = (\frac{1}{2},\cdots,\frac{1}{2})$.
This lattice has coordination number $k_{bcc(d)}=2^d$.  Using eq. (\ref{zint}),
Ref. \cite{sw} obtained
\beq 
z_{bcc(d)} = d \ \ln 2 + I_{bcc(d)}
\label{zbccd}
\eeq
where
\beqs
I_{bcc(d)} & = &  \frac{1}{2} \int_{-\pi}^\pi 
\biggl [ \prod_{j=1}^d {d\theta_j \over {2\pi}} \biggr ]
\ln \biggl ( 1 - \prod_{j=1}^d \cos^2(\theta_j/2) 
\biggr ) \cr\cr
&=& 
\int_{-\pi}^\pi \biggl [ \prod_{j=1}^d {d\theta_j \over {2\pi}} \biggr ]
\ln \biggl ( 1 - \prod_{j=1}^d \cos\theta_j \biggr ) \ . 
\label{ibccd}
\eeqs
Expanding the logarithm and carrying out the integration term by term yields
the infinite series representation \cite{sw}
\beq
I_{bcc(d)} = -\frac{1}{2}\sum_{\ell=1}^\infty \frac{1}{\ell}
\Biggl (\frac{(2\ell)!}{2^{2\ell} \, (\ell !)^2} \Biggr )^d
\label{ibccdseries}
\eeq
We now sum this series exactly. First, 
\beqs
I_{bcc(d)}
& = & -\frac{1}{2}\sum_{\ell=1}^\infty \frac{(\ell-1)! \, [(2\ell)!]^d}
{2^{2\ell d} \, (\ell !)^{2d+1}} \cr\cr
& = & -\frac{1}{2} \sum_{k=0}^\infty \frac{(k!)^2 \, [(2k+2)!]^d}
{2^{2(k+1)d} \, [(k+1)!]^{2d+1} \, k!} \cr\cr
& = & -\frac{1}{2^{d+1}}\sum_{k=0}^\infty 
\frac{[\Gamma(k+1)]^2 \, [\Gamma(2k+3)]^d}
     {2^{(2k+1)d} \, [\Gamma(k+2)]^{2d+1} \, k!} \ . 
\label{ibccdser1}
\eeqs
Next, we use the duplication formula for the Euler gamma function, 
\beq
\Gamma(2z) = (2\pi)^{-1/2} \, 2^{2z-\frac{1}{2}} \, \Gamma(z) \, 
\Gamma(z + \frac{1}{2}) 
\label{dup}
\eeq
with $z=k+\frac{3}{2}$, together with $\Gamma(1/2)=\sqrt{\pi}$, to express
\beq
\frac{\Gamma(2k+3)}{2^{2k+1} \, \Gamma(k+2)} = 
\frac{\Gamma(k+ \frac{3}{2})}{\Gamma(\frac{3}{2})} \ . 
\label{dupap}
\eeq
Substituting this into eq. (\ref{ibccdser1}), we have
\beqs
I_{bcc(d)} 
& = & -\frac{1}{2^{d+1}}\sum_{k=0}^\infty 
\frac{[\Gamma(k+1)]^2 \, [\Gamma(k+\frac{3}{2})/\Gamma(\frac{3}{2})]^d}
     {[\Gamma(k+2)]^{d+1} \, k!} \cr\cr
& = & -2^{-(d+1)} \, {}_{d+2}F_{d+1}([1,1,3/2,\cdots,3/2],
[2,\cdots,2],1)
\label{ibccdfinal}
\eeqs
where there are $d+2$ entries in first square bracket $[\cdots]$ and $d+1$ 
entries in the second square bracket $[\cdots]$ in the argument, and 
${}_{p}F_{q}$ is the generalized hypergeometric function, 
\beq
{}_p F_q ([a_1,\cdots,a_p],[b_1,\cdots,b_q],x) = \sum_{k=0}^\infty 
\Biggl ( \frac{\prod_{j=1}^p (a_j)_k}
     {\prod_{r=1}^q (b_r)_k} \Biggr ) \frac{x^k}{k!}
\label{fpq}
\eeq
where $c_n = \Gamma(c+n)/\Gamma(c)$. Hence, 
\beq
z_{bcc(d)} = d \, \ln 2 - 2^{-(d+1)} \ 
{}_{d+2}F_{d+1}([1,1,3/2,\cdots,3/2], [2,\cdots,2],1) \ . 
\label{zbccdfinal}
\eeq

We comment on some special cases.  For $d=1$, the $bcc(1)$ lattice with free
(periodic) boundary conditions degenerates effectively to a line (circuit)
graph, for which, respectively, $N_{ST}=1$ and $N_{ST}=n$; in both cases, it
follows that $z_{bcc(1)}=0$.  Using the value
${}_3F_2([1,1,3/2],[2,2],1)=4\ln2$, we recover this elementary result.  For
$d=2$, the $bcc(2)$ lattice is equivalent to the square lattice, for which
$z_{sq} = (4/\pi)\beta(2)=1.1662436..$ \cite{temperley,wu77} , where $\beta(s)
= \sum_{n=0}^\infty (-1)^n (2n+1)^{-s}$ and $\beta(2)=C=0.915965594177..$ is
the Catalan constant. The general result (\ref{zbccd}) with (\ref{ibccdfinal})
evaluated for $d=2$ agrees with this, since ${}_4F_3([1,1,3/2,3/2],[2,2,2],1)=
16(\ln 2 - (2C/\pi))$.  Our general exact result for $z_{bcc(d)}$ provides
quite accurate values for higher values of $d$, which we list in Table
\ref{zbccvalues}, together with the corresponding ratios (\ref{rupper}) which
give a comparison with the upper bound (\ref{mcybound}).  Evidently, the exact
values are very close to this upper bound and move closer as $d$ increases.

\begin{table}
\caption{\label{zbccvalues} Values of $z_{bcc(d)}$ and $r_{bcc(d)}$.}
\footnotesize\rm
\begin{tabular*}{\textwidth}{@{}l*{15}{@{\extracolsep{0pt plus12pt}}l}}
\br
$d$ & $z_{bcc(d)}$ & $r_{bcc(d)}$ \\
\mr
1  &  0                  &              $-$         \\ 

2  &  1.166243616123275  &    0.9587702228064145    \\

3  &  1.990191418271941  &    0.9912457055306051    \\

4  &  2.732957535477362  &    0.9977098978275579    \\

5  &  3.447331914522398  &    0.9993413280070963    \\ 

6  &  4.150116933352462  &    0.9998002121159708    \\

7  &  4.847789269805724  &    0.9999373061649456    \\

8  &  5.543104959793989  &    0.9999798500846987    \\ 

9  &  6.237305017795394  &    0.9999934053622532    \\
 
10 &  6.930967870288660  &    0.9999978103135475    \\
\br
\end{tabular*}
\end{table}

\section{fcc Lattice} 

The face-centered cubic (fcc) lattice has coordination number $k_{fcc}=12$
and a unit cell consisting of the $\nu_{fcc}=4$ 
vertices $(0,0,0)$, $(0,\frac{1}{2},\frac{1}{2})$,
$(\frac{1}{2},0,\frac{1}{2})$, and $(\frac{1}{2},\frac{1}{2},0)$.  For this
lattice, $M(\theta_1,\theta_2,\theta_3)$ is \cite{sw}
\beq
M(\theta_1,\theta_2,\theta_3) =
\left(\begin{array}{cccc}
12       &   -(v_2v_3)^*   &   -(v_1v_3)^*   &   -(v_1v_2)^*  \\
-v_2v_3  &      12         &   -v_1^*v_2     &   -v_1^*v_3    \\
-v_1v_3  &   -v_1v_2^*     &       12        &   -v_2^*v_3    \\
-v_1v_2  &   -v_1v_3^*     &    -v_2v_3^*    &       12  \end{array} \right) 
\label{Mfcc}
\eeq
where $v_j=1+e^{i\theta_j}, j=1,2,3$. The evaluation of the determinant yields
\beq
z_{fcc} = \ln (12) +\frac{1}{4} \int_{-\pi}^{\pi}{{d\theta_1}\over{2\pi}}
 \int_{-\pi}^{\pi}{{d\theta_2}\over{2\pi}}
 \int_{-\pi}^{\pi}{{d\theta_3}\over{2\pi}}
 \ln  F(\theta_1,\theta_2,\theta_3) 
\label{zfcc}
\eeq
where, with the abbreviation $c_j \equiv \cos(\theta_j/2)$, 
\beqs
F(\theta_1,\theta_2,\theta_3) 
 & = & \Bigl [ 1 + \frac{1}{3}(-c_2 c_3 + c_3 c_1 + c_1 c_2) \Bigr ]
       \Bigl [ 1 + \frac{1}{3}( c_2 c_3 - c_3 c_1 + c_1 c_2) \Bigr ] \cr\cr
  & \times &
       \Bigl [ 1 + \frac{1}{3}(c_2 c_3 + c_3 c_1 - c_1 c_2) \Bigr ]
       \Bigl [ 1 - \frac{1}{3}( c_2 c_3 + c_3 c_1 + c_1 c_2) \Bigr ] \cr\cr
& = & 1 - \frac{2}{9}[(c_1 c_2)^2 + (c_2 c_3)^2 + (c_3 c_1)^2]
        - \frac{8}{27}(c_1 c_2 c_3)^2 \cr\cr
       & - & \frac{2}{81}(c_1 c_2 c_3)^2 (c_1^2 + c_2^2 + c_3^2)
        + \frac{1}{81}[(c_1c_2)^4 + (c_2 c_3)^4 + (c_3 c_1)^4] \ . \cr\cr
& & 
\label{Ffcc}
\eeqs
(This corrects an algebraic error in eq. (5.3.3) of Ref. \cite{sw}).
Evaluating this numerically, we find that $z_{fcc} \simeq 2.41292$.
Substituting $z_{fcc}$ into eq. (\ref{rupper}), we get $r_{fcc} \simeq
0.98915$, so that the upper bound (\ref{rupper}) is very close to the actual
value of the growth constant.

\section{$4 \cdot 8 \cdot 8$ Lattice}

An Archimedean lattice is a uniform tiling of the plane by regular polygons in
which all vertices are equivalent.  Such a lattice can be defined by the
ordered sequence of polygons that one traverses in making a complete circuit
around the local neighborhood of any vertex.  This is indicated by the notation
$\Lambda = (\prod_i p_i^{a_i})$, meaning that in this circuit, a regular
$p_i$-sided polygon occurs contiguously $a_i$ times.  We consider here the $4
\cdot 8 \cdot 8$ lattice involving the tiling of the plane by squares and
octagons.  In eq. (4.11) of Ref. \cite{sw}, the asymptotic growth constant for
this lattice was calculated to be 
\beqs
z_{488} & = &  {1\over 2} \ln 2 + {1\over {4} } \int_{-\pi}^{\pi}
{{d\theta_1}\over {2\pi}}\int_{-\pi}^{\pi} {{d\theta_2}\over {2\pi}}
\ln \Big[7-3(\cos\theta_1+\cos\theta_2) - \cos\theta_1\cos\theta_2 \Big]
\cr\cr
&=& {1\over 4} \ln 2 +{1\over {4\pi} } \int_{0}^{\pi}d\theta
 \ln\Big[7-3\cos\theta + 4\sin(\theta/2) \sqrt{5-\cos\theta} \ \Big] 
\label{z488int}
\eeqs
where the integral on the second line of eq. (\ref{z488int}) is obtained by
doing one of the two integrations in the expression on the first line. These
integrals were evaluated numerically to obtain the result $z_{488} =
0.786684(1)$, where the number in parentheses indicates the estimated error in
the last digit.

We have derived an exact closed form expression for this integral.  We begin by
recasting the integral in the equivalent form.  
\beqs
z_{488} & = & \frac{1}{4}\ln 2 + \frac{1}{2\pi}\int_0^\pi d\theta \
\ln \Bigl ( 2\sin(\theta/2)+\sqrt{4+2\sin^2(\theta/2)} \ \Bigr ) \cr\cr
& = & \frac{3}{4}\ln 2 + \frac{1}{\pi}\int_0^{\pi/2} d\phi \
\ln \Bigl ( \sin(\phi) + \sqrt{1+(1/2)\sin^2(\phi)} \ \Bigr ) \ .
\label{int1}
\eeqs
That is, 
\beq
z_{488} = \frac{3}{4}\ln 2 + I(1/\sqrt{2} \, )
\label{z488i}
\eeq
where
\beq
I(a) = \frac{1}{\pi} \int_0^{\pi/2} d\phi \ \ln \Bigl ( \sin\phi 
+ \sqrt{1+a^2 \sin^2\phi } \ \Bigr ) \ . 
\label{iint}
\eeq
In eq. (\ref{iint}), with no loss of generality, we take $a$ to be nonnegative.
We will give a general result for $I(a)$ and then specialize to our case
$a=1/\sqrt{2}$.  First, we note that $I(1)=C/\pi$, where $C$ is the Catalan
constant. Next, assume $0 \le a < 1$. Taking the derivative with respect to $a$
and doing the integral over $\phi$ in eq. (\ref{iint}), we get
\beq
I'(a) = \frac{-a/2 + (2/\pi)\tan^{-1}a}{(1-a^2)} \ . 
\label{iprime}
\eeq
To calculate $I(a)$, we then use $I(a)-I(0) = \int_{0}^{a} I'(x) dx$ and
observe that 
\beq
I(0) = \frac{1}{\pi} \int_{0}^{\pi/2} d\phi \ \ln (\sin(\phi)+1) =
-\frac{\ln 2}{2} + \frac{2C}{\pi}
\label{izero}
\eeq
We also make use of the integrals 
\beq
\int_{0}^{a} \frac{x}{(1-x^2)} \ dx = -\frac{1}{2} \ln (1-a^2)
\label{int1a}
\eeq
and
\beq
\int_{0}^{a} \frac{\tan^{-1}x}{(1-x^2)} \ dx = -\frac{C}{2} 
-\frac{\pi}{8} \ln \biggl ( \frac{1+a}{1-a} \biggr ) + 
\frac{1}{2} {\rm Ti}_2 \biggl ( \frac{1+a}{1-a} \biggr )
\eeq
to obtain
\beq
I(a) = \frac{C}{\pi} + \frac{1}{2} \ln \Big ( \frac{1-a}{2} \Big )
+ \frac{1}{\pi}{\rm Ti}_2 \biggl ( \frac{1+a}{1-a} \biggr ) \quad {\rm if} \ \
0 \le a < 1, 
\label{iabelow1}
\eeq
where Ti$_2(x)$ is the inverse tangent integral \cite{invtan}, 
\beqs
{\rm Ti}_2(x) & = & \int_0^x \frac{\tan^{-1} y}{y} \ dy \cr\cr
        & = & x[ \ {}_3F_2([1,1/2,1/2],[3/2,3/2],-x^2) \ ] \ . 
\label{ti2}
\eeqs
(Here the arctangent is taken to lie in the range $-\pi/2 < \tan^{-1} y <
\pi/2$.)  Evaluating our result (\ref{iabelow1}) for $I(a)$ at
$a=1/\sqrt{2}$ and substituting into eq. (\ref{z488i}), we obtain the exact,
closed-form expression
\beq
z_{488} = \frac{C}{\pi} + \frac{1}{2} \ln (\sqrt{2} \, - 1)
+ \frac{1}{\pi} {\rm Ti}_2(3+2\sqrt{2} \, ) \ . 
\label{z488an}
\eeq
The numerical evaluation of eq. (\ref{z488an}) agrees with the evaluation given
in Ref. \cite{sw} to the accuracy quoted there and allows one to obtain higher
accuracy; for example, to 15 significant figures, $z_{488} =
0.786684275378832$.  We note that the ${\rm Ti}_2$ function also appears at
intermediate stages in the derivation of $z_{tri}$ for the triangular lattice
\cite{gw}.  For completeness, we have also calculated $I(a)$ for $a > 1$ with
the result
\beq 
I(a) = \frac{C}{\pi} + \frac{1}{2} \ln \bigg [ \frac{(a+1)^2}{2(a-1)} \bigg ]
+ \frac{1}{\pi}{\rm Ti}_2 \biggl ( \frac{1+a}{1-a} \biggr ) \quad {\rm if} \ \
 a > 1 \ . 
\label{iaabove1}
\eeq

\section{Acknowledgments}
S.-C.C. thanks C.-H. Chen for use of computer facilities and software.
R. S. thanks Prof. F. Y. Wu for the collaboration on Ref. \cite{sw}. This
research was partially supported by the Taiwan NSC grant NSC-94-2112-M-006-013
(S.-C.C.) and the U.S. NSF grant PHY-00-98527 (R.S.).

\end{document}